\documentclass{aa}
\usepackage{txfonts,psfig,epsfig}
\begin{document}

\parindent0cm

\title{Evidence for grain growth in T\,Tauri disks\thanks{Based on
    observations made with the ESO 3.6m Telescope at the La Silla Observatory
    under program ID 68.D-0537(A), 69.C-0268 and 70.C-0544.}}


\author{F. Przygodda \inst{1}
          \and
          R. van Boekel \inst{2}
          \and
          P. \`Abrah\`am \inst{3}
          \and
          S. Y. Melnikov \inst{4}
          \and
          L. B. F. M Waters \inst{2,5}
          \and
          Ch. Leinert \inst{1}}

\offprints{F. Przygodda}

\institute{Max-Planck-Institut f\"ur Astronomie, K\"onigstuhl 17, 69117
  Heidelberg, Germany, email: przygodda@mpia.de
  \and
  Astronomical Institute ``Anton Pannekoek'', University of Amsterdam, Kruislaan 403, 1098 SJ Amsterdam, The Netherlands
  \and
  Konkoly Observatory of the Hungarian Academy of Sciences, H-1525 Budapest, 
  P.O. Box 67, Hungary
  \and
  Ulugh Beg Astronomical Institute, Academy of Sciences of Uzbekistan,
  Astronomical str. 33, Tashkent 700051, Uzbekistan
  \and
  Instituut voor Sterrenkunde, Katholieke Universiteit Leuven, Celestijnenlaan 200B,
  B-3001 Heverlee, Belgium}

\date{Received 10 September 2003 / Accepted 6 November 2003}

\abstract{In this article we present the results from mid-infrared
  spectroscopy of a sample of 14 T\,Tauri stars with silicate emission. The
  qualitative analysis of the spectra reveals a correlation between the
  strength of the silicate feature and its shape similar to the one which was
  found recently for the more massive Herbig Ae/Be stars by van Boekel~et~al.
  (2003). The comparison with theoretical spectra of amorphous olivine
  ([Mg,Fe]$_2$SiO$_4$) with different grain sizes suggests that this
  correlation is indicating grain growth in the disks of T\,Tauri stars.
  Similar mechanisms of grain processing appear to be effective in both groups
  of young stars.

 }

\maketitle

%

\section{Introduction}

T\,Tauri stars are low mass young stellar objects with ages of
$10^6-10^7$\,yr. One of the characteristics of so-called classical T\,Tauri
stars is a distinctive IR excess. The origin of the IR radiation is the
accretion disk which plays a crucial role in star formation by transporting
material to the central star while angular momentum is transferred outwards.
Different models were proposed to bring the observed facts into a general
framework (\cite{pringle74}; \cite{adams87}; \cite{chiang97}). Yet there
remain many open questions concerning the details of configuration, chemistry
and evolution of disks. A particular feature of the infrared spectra of such
disks is the silicate band in the region from 8 to 12\,$\mu$m.  The band
originates from the stretching mode of the Si-O bond of silicate minerals like
olivine ([Mg,Fe]$_2$SiO$_4$), forsterite (Mg$_2$SiO$_4$), enstatite
(MgSiO$_3$) and silica (SiO$_2$).  The silicate feature observed in T\,Tauri
stars appears in emission as well as in absorption (\cite{cohen85}). Silicate
emission is assumed to emerge from a warm, optically thin disk layer (disk
atmosphere) which is heated by the radiation of the central star
(\cite{chiang97}, \cite{natta00}). Honda~et~al. (2003) showed that crystalline
silicates are present in T Tauri stars, indicating substantial grain
processing. In this Paper we present studies on the silicate emission from a
sample of 14 T\,Tauri stars based on mid-infrared spectroscopy performed
during three observation campaigns in February, June and December 2002 on the
ESO 3.6\,m telescope at La Silla. We analyse the spectra in particular with
respect to a possible correlation between shape and strength of the silicate
feature. Such a correlation has been recently observed for Herbig Ae/Be stars
(\cite{boekel03}) and has been interpreted as evidence of grain processing in
the circumstellar disks of those stars.

\section{Observations and data reduction}
\label{redu}

The observations of the T\,Tauri stars were performed with TIMMI2, the Thermal
Infrared Multi Mode Instrument 2 (\cite{reimann98}) installed at the 3.6 m
telescope at ESO's observatory at La Silla, Chile.  We observed each object in
two modes: first we used the imaging mode to obtain photometry at 11.9\,$\mu$m
(band width 1.2\,$\mu$m). In a number of cases in addition we used the
8.9\,$\mu$m filter (band width 0.8\,$\mu$m).  Second we performed spectroscopy
using the low-resolution grism mode (R=160) with a slit width of 1.2\arcsec.
In both modes the system was working with chopping and nodding with an
amplitude of 10\arcsec.  In our data reduction, the measurements of the
spectrophotometric standards were used to determine the shape of the spectra,
and the photometric data at 11.9\,$\mu$m were used to establish the absolute
flux level.  The spectrum of the atmospheric extinction, which is needed to
perform the first of these steps, was determined from observations of standard
stars at different airmass. We regularly observed standard stars to monitor
variations in atmospheric transmission.  The flux data for the
spectrophotometric standard stars, (based on models by \cite{cohen98}), were
taken from ESO's TIMMI2 webpage.

In total we observed 21 T\,Tauri stars with the silicate feature in emission.
We selected the 14 objects for which the silicate feature was measured with a
signal-to-noise ratio better than 2 (see Tab.\,\ref{tab1}).  Several of these
are known to be binary or multiple systems. In two cases we were able to
obtain separate spectra of the components.  The remaining 7 sources (DO\,Tau,
DQ\,Tau, FU\,Ori, BBW\,76, VZ\,Cha, VW\,Cha, Sz\,82) were degraded by noise to
the extent that a meaningful spectral analysis was not possible.  For some
stars the observations were repeated to verify the results.

\begin{table}[t]
\centering
\caption{List of observed sources with the silicate feature in emission.}
\label{tab1}
\begin{tabular}{lcccc}
\hline
\hline
Object & Flux at           & Observed               &  Multiple & ISOPHOT\\
       & 11.9\,$\mu$m & at Run $^{\mathrm{a}}$ & Source $^{\mathrm{b}}$ 
                                                           &     spectrum\\
       &      (Jy)      &                        &           &  available\\
\hline
AK Sco       &  2.1\, & B      & * & * \\
AS 205       &  8.4 & B      & * & \\
CR Cha       &  0.8 & C      &   & * \\
DR Tau       &  2.0 & C      &   & * \\
GG Tau       &  0.8 & C      & * & \\
GW Ori       &  6.3 & C      & * &  \\
Glass I      &  8.8 & A,B,C  & * & *\\
Haro 1-16    &  0.9 & B      &   & \\
HBC 639      &  2.3 & A      &   & \\
RU Lup       &  2.4 & B      &   & \\
S CrA        &  5.1 & B      & * &  *\\
SU Aur       &  3.6 & C      & * & \\
TW Hya       &  0.6 & C      &   & \\
WW Cha       &  5.3 & B,C    &   & * \\
\hline
\end{tabular}
\begin{list}{}{}
\item[$^{\mathrm{a}}$]A: 5.+6. Feb. 2002, B: 7.+8. Jun. 2002, C:
  24.-27. Dec. 2002
\item[$^{\mathrm{b}}$]for AS 205 and S CrA we were able to 
  obtain separate spectra of the components
\end{list}
\vspace*{-0.5cm}
\end{table}

\begin{figure}[t]
\begin{center}
\epsfig{file=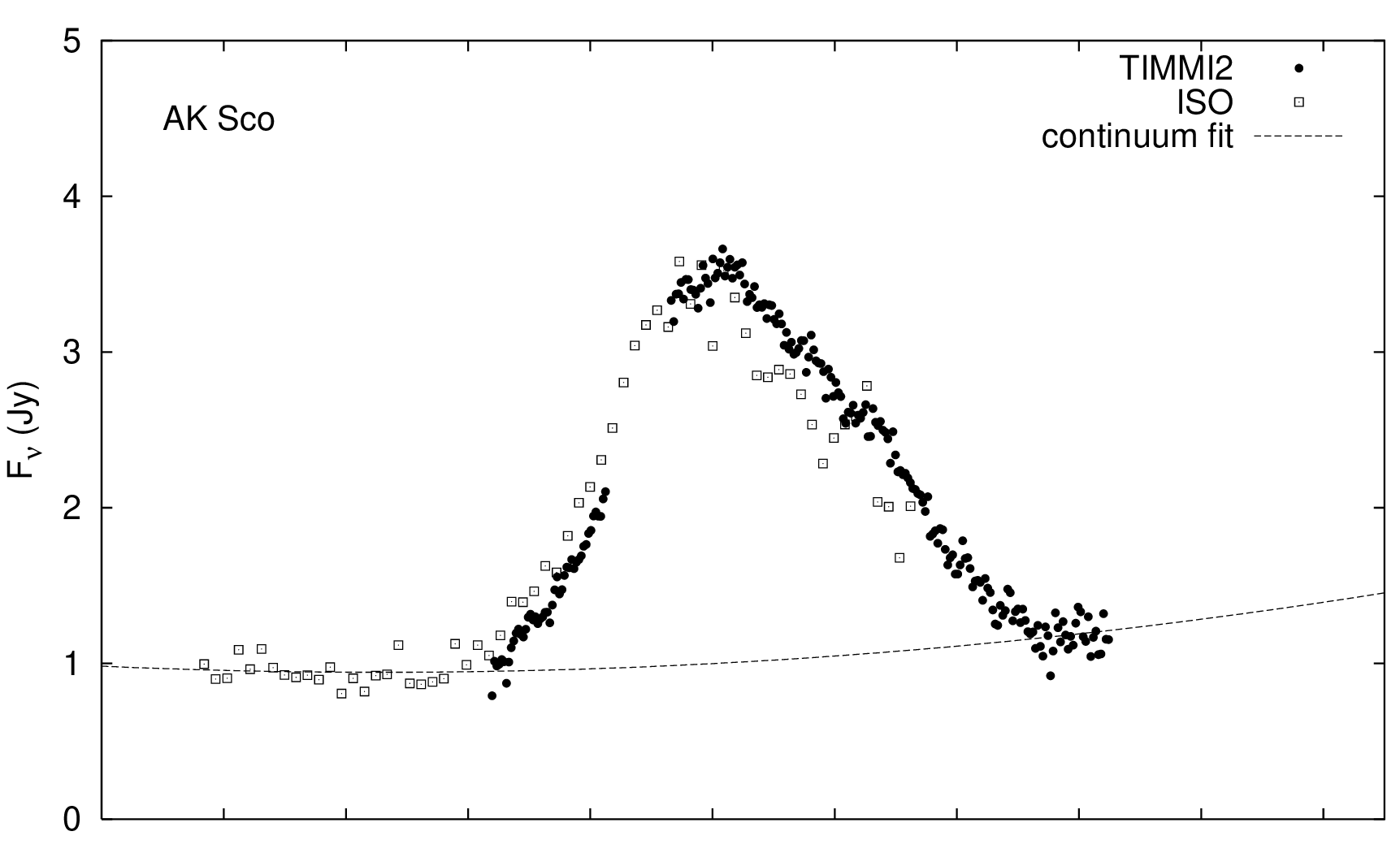,angle=270,width=7.5cm}
\epsfig{file=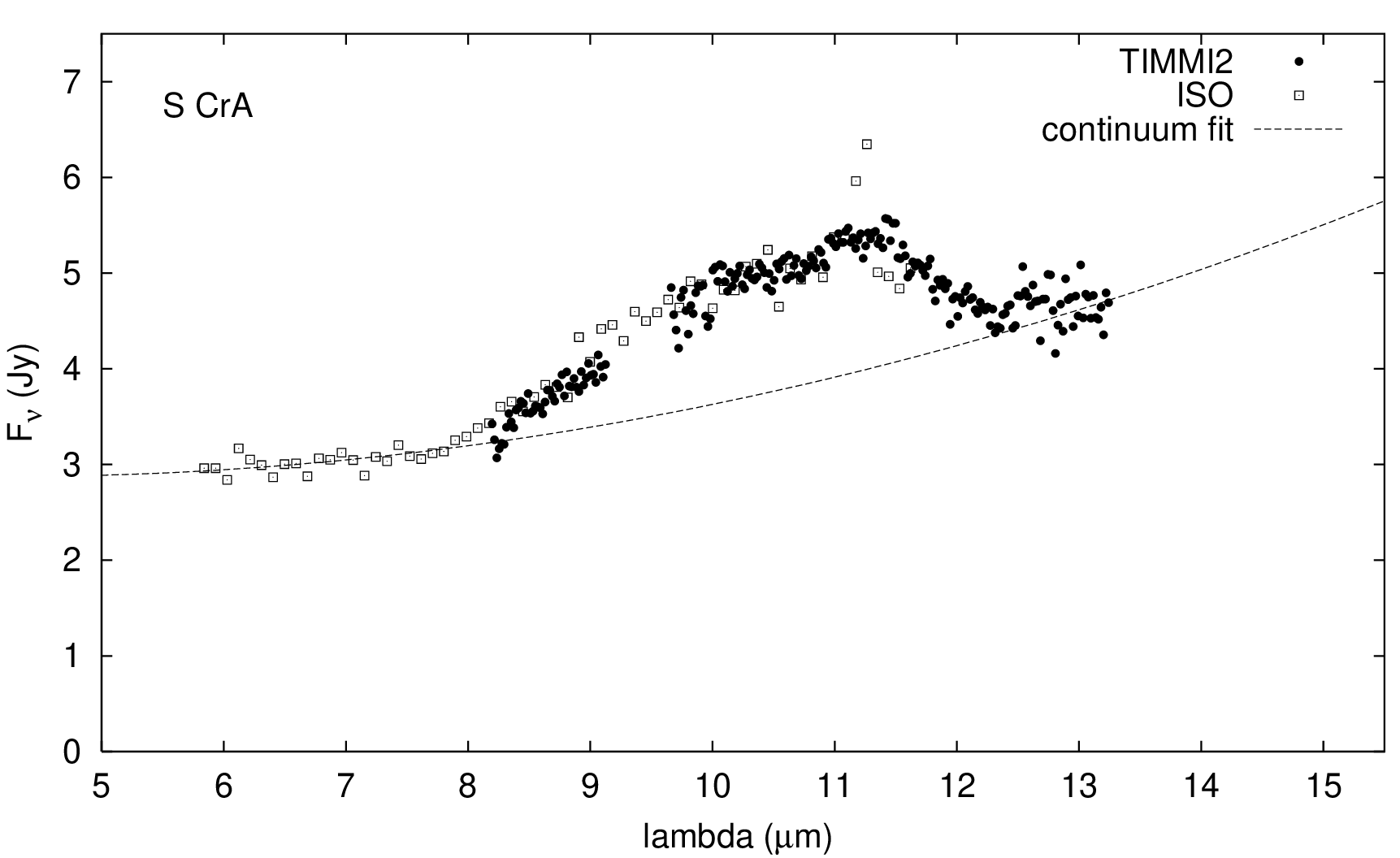,angle=270,width=7.5cm}
\end{center}
\vspace*{-0.4cm}
\caption{Comparison of TIMMI2 and ISOPHOT data for two objects. The
  combined data were used to estimate the continuum by fitting a second order
  polynomial to the data points left of 8 \,$\mu$m and longwards of
  12.8\,$\mu$m.}
\label{plot12}
\end{figure}

For our studies we assumed that the continuum outside of the silicate band is
actually reached at the edges of our measured spectral range (at 8.2\,$\mu$m
and at 13.0\,$\mu$m). This assumption is supported by comparison of our
spectra with ISOPHOT spectra (\'Abrah\'am et. al 2003 in prep, see also
\cite{natta00}). Figure \ref{plot12} illustrates this for two examples.  Here,
the continuum was estimated by fitting a second order polynomial to all data
points shortwards of 8\,$\mu$m and beyond 12.8\,$\mu$m. Note that the fitted
curve is matching the ends of the TIMMI2 spectra and has an almost linear
characteristic over the range of the silicate feature. Since we were not able
to find such additional data of good quality for all objects of our sample, we
estimated the continuum level within the silicate feature by a linear
interpolation between the end points of the TIMMI2 spectra at 8.2\,$\mu$m and
at 13.0\,$\mu$m, where the values for these end points were obtained by
averaging over the data over an interval of 0.2\,$\mu$m. The continuum level
determined this way agreed with the second order fit, where available, to
typically $\pm$\,10\,\% .  For all sources we consistently used the
derived linearly varying continuum level to determine the continuum
normalized\footnote{to preserve the shape of the emission feature we define:\\
  $ F_{\mbox{\tiny norm}}(\lambda) =
  1+[F_{\mbox{\tiny total}}(\lambda)-F_{\mbox{\tiny continuum}}(\lambda)]/
  \langle F_{\mbox{\tiny continuum}} \rangle$} spectra; these are shown
in Fig.\,\ref{plot3}.

\section{Analysis of the silicate feature}

\begin{figure*}
\epsfig{file=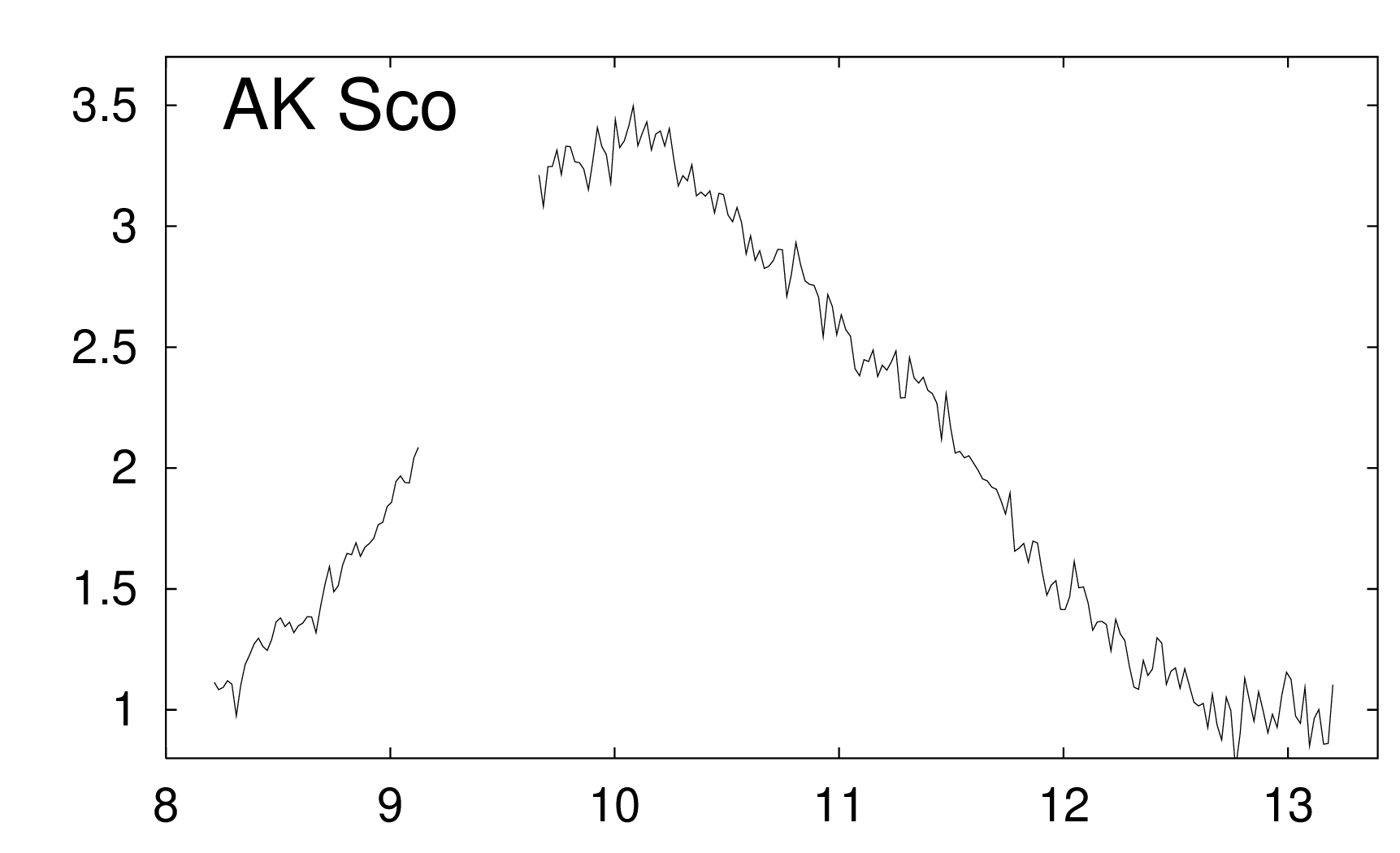,angle=270,width=4.45cm}
\epsfig{file=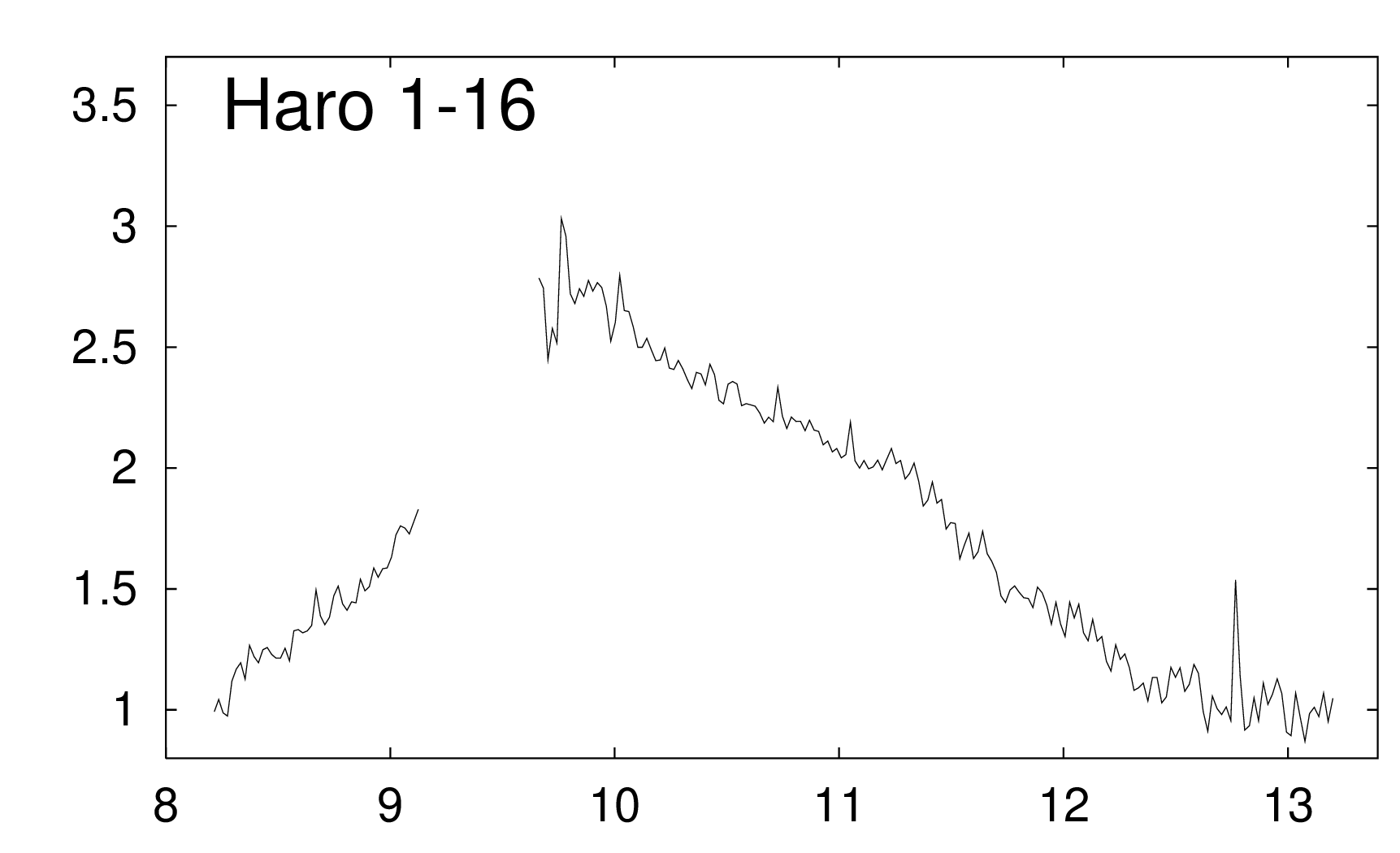,angle=270,width=4.45cm}
\epsfig{file=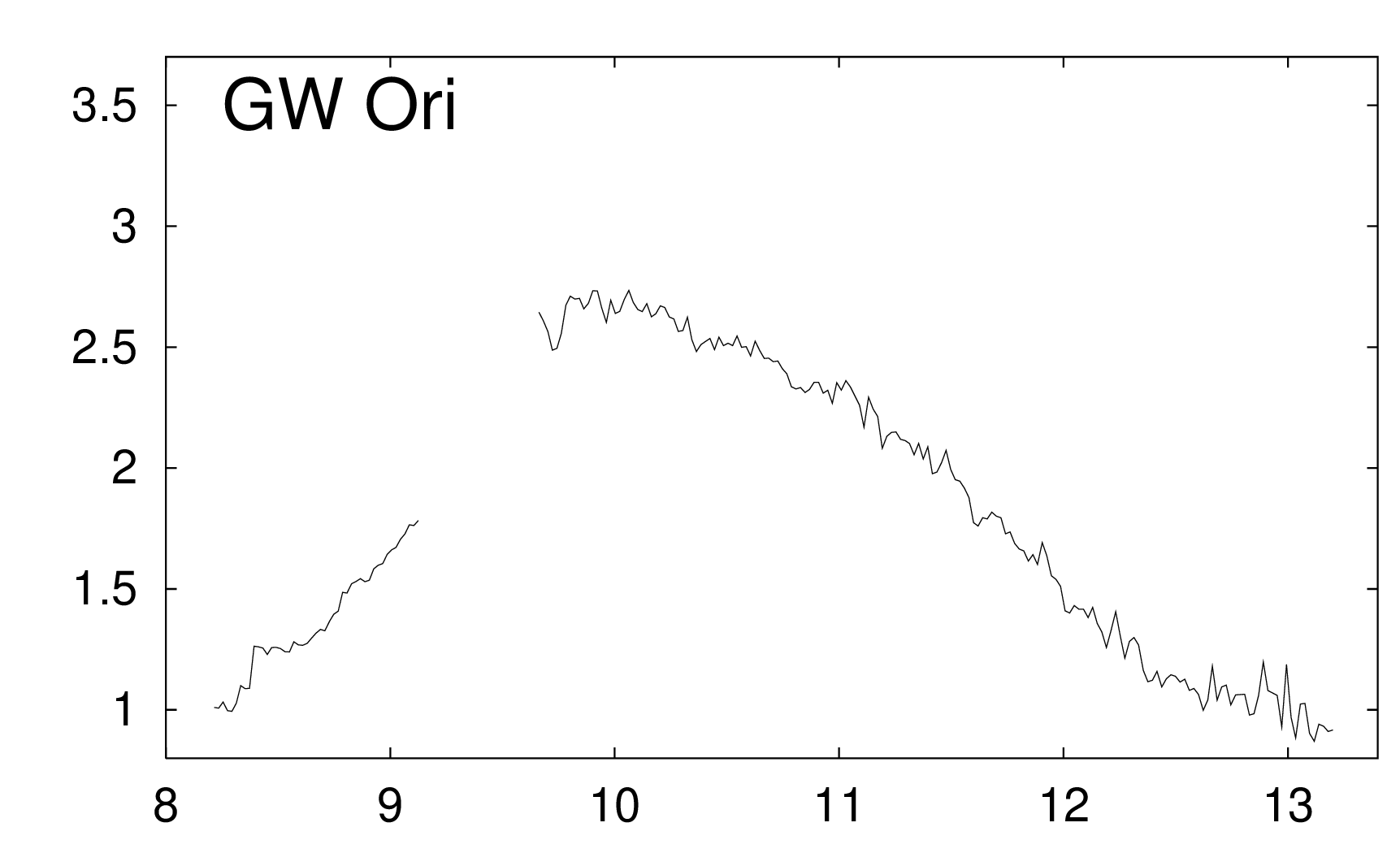,angle=270,width=4.45cm}
\epsfig{file=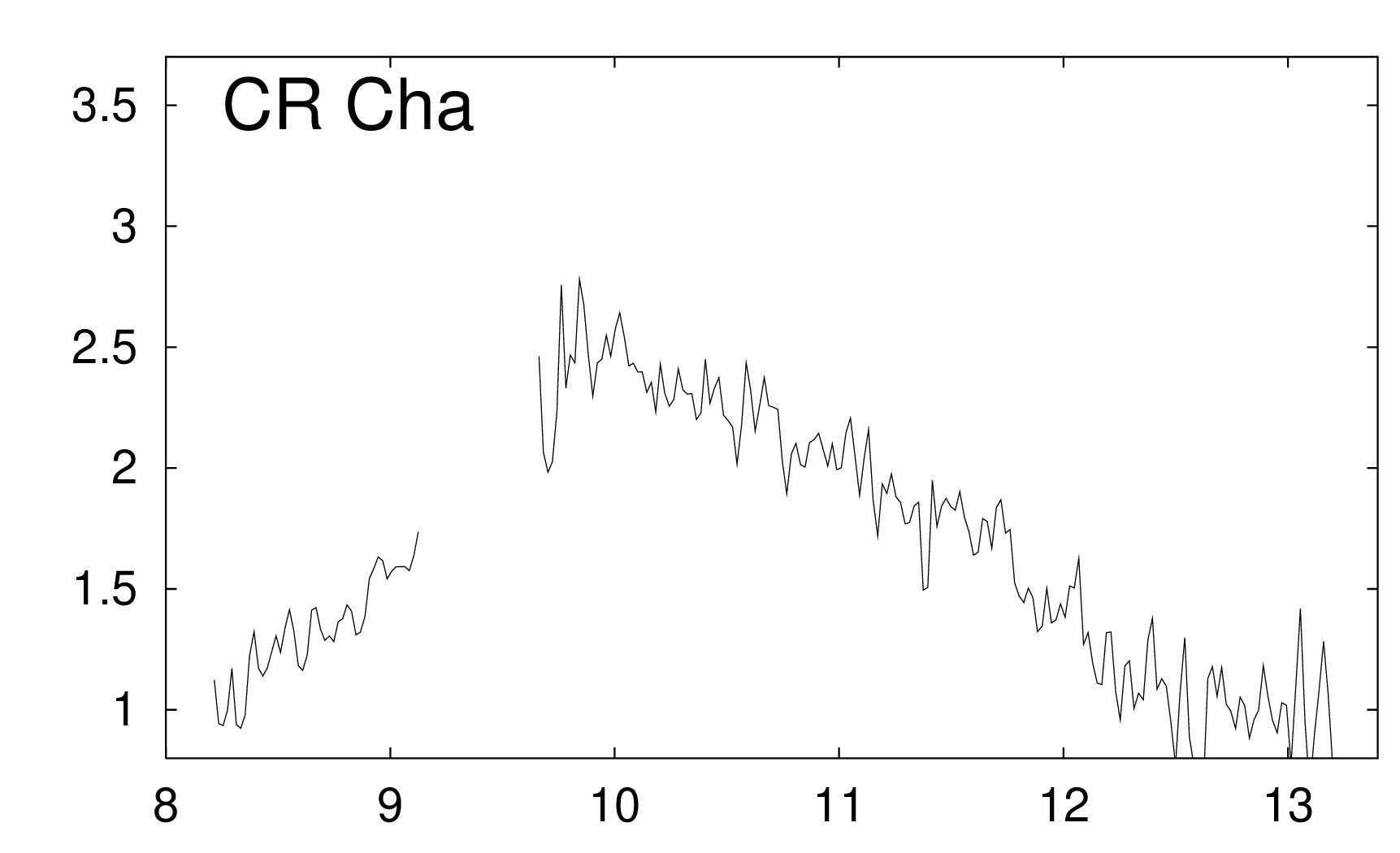,angle=270,width=4.45cm}
\epsfig{file=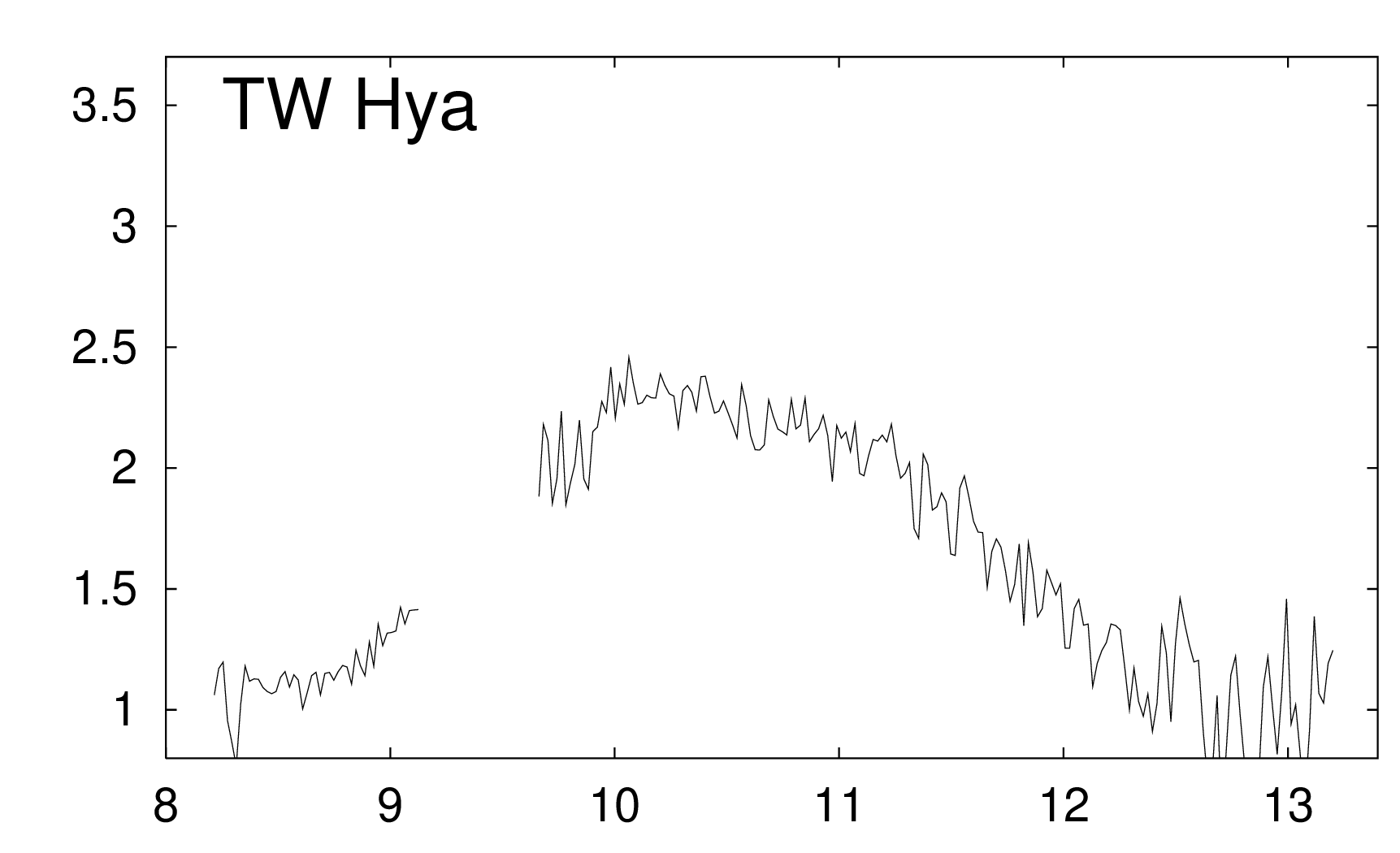,angle=270,width=4.45cm}
\epsfig{file=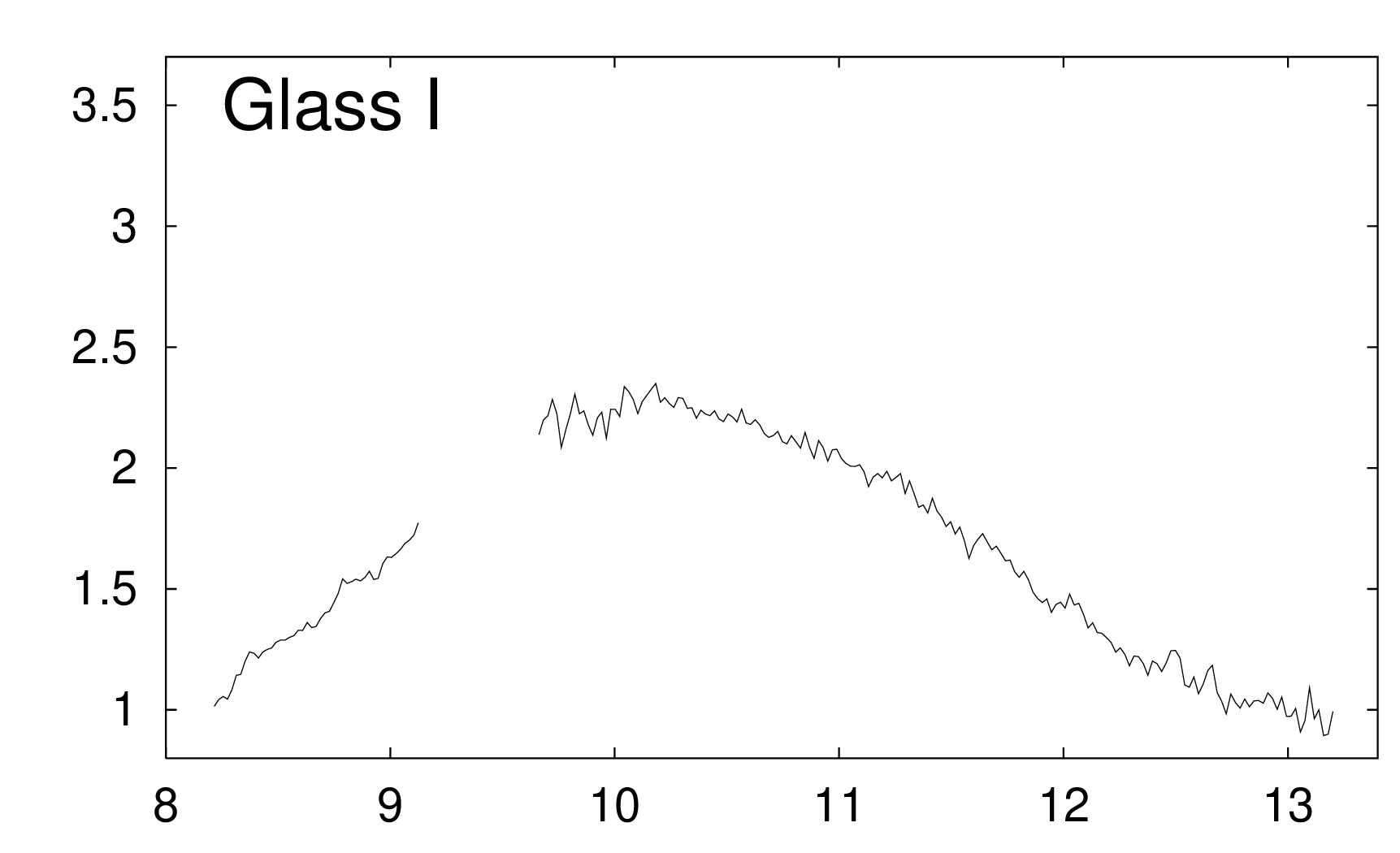,angle=270,width=4.45cm}
\epsfig{file=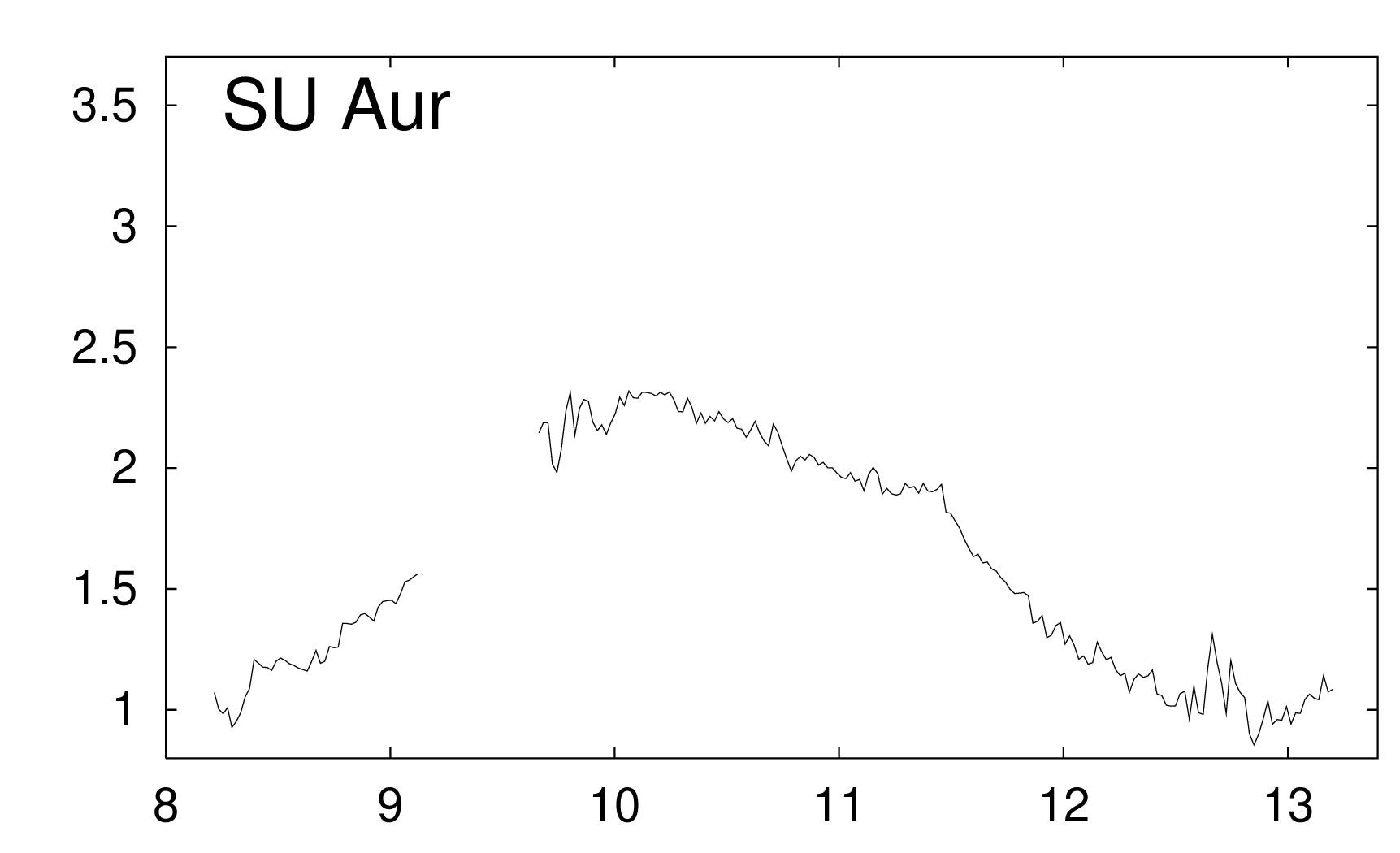,angle=270,width=4.45cm}
\epsfig{file=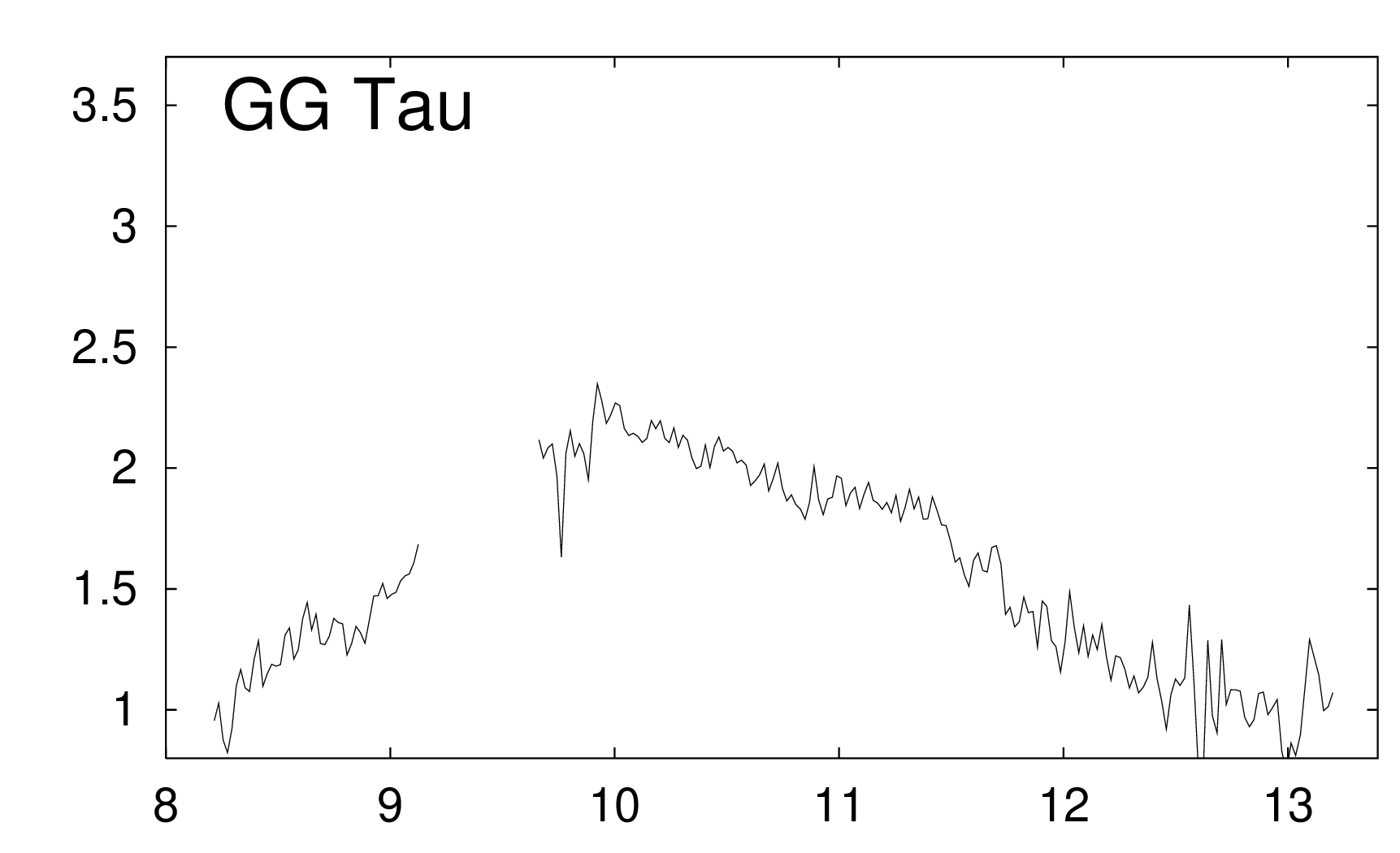,angle=270,width=4.45cm}
\epsfig{file=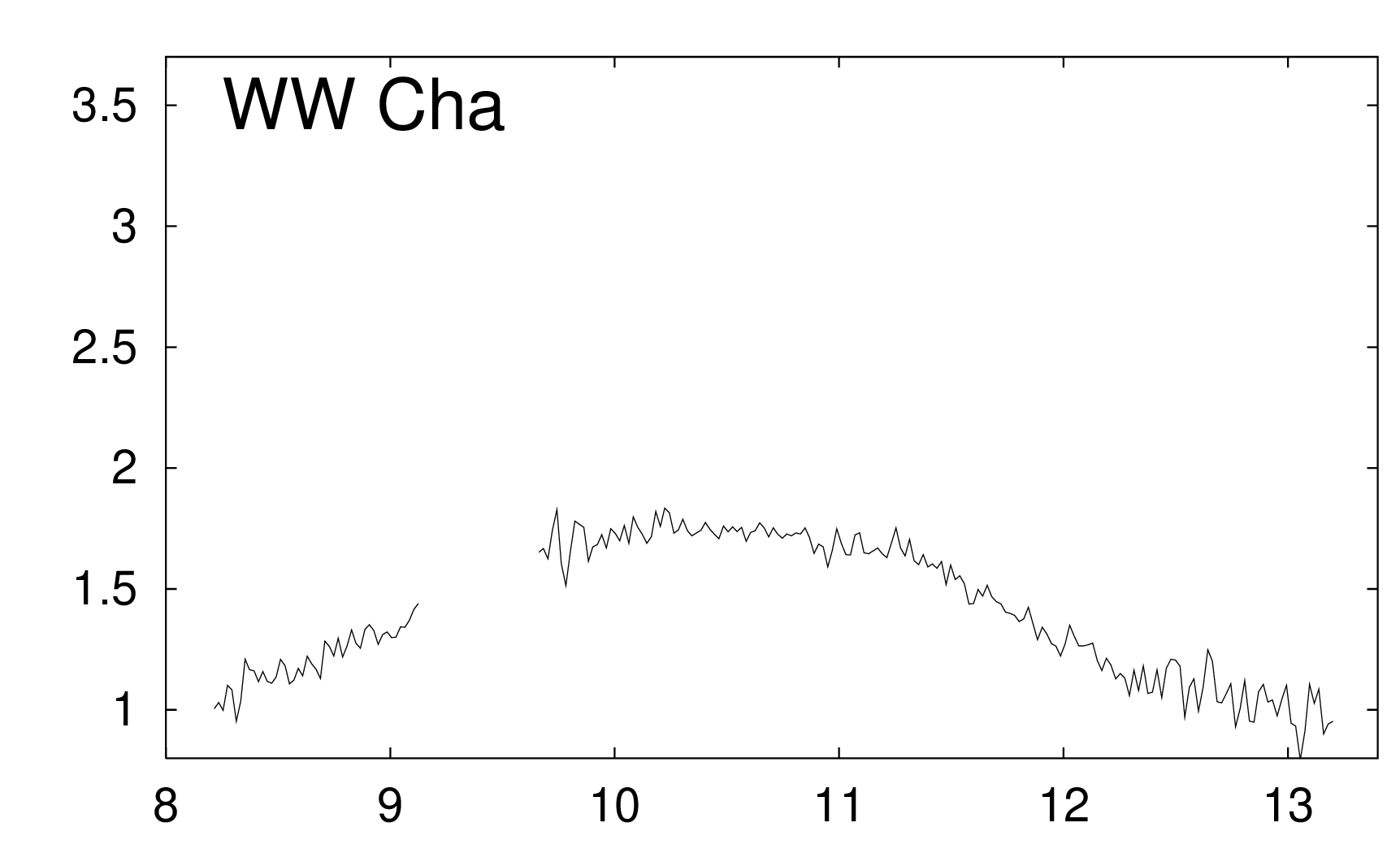,angle=270,width=4.45cm}
\epsfig{file=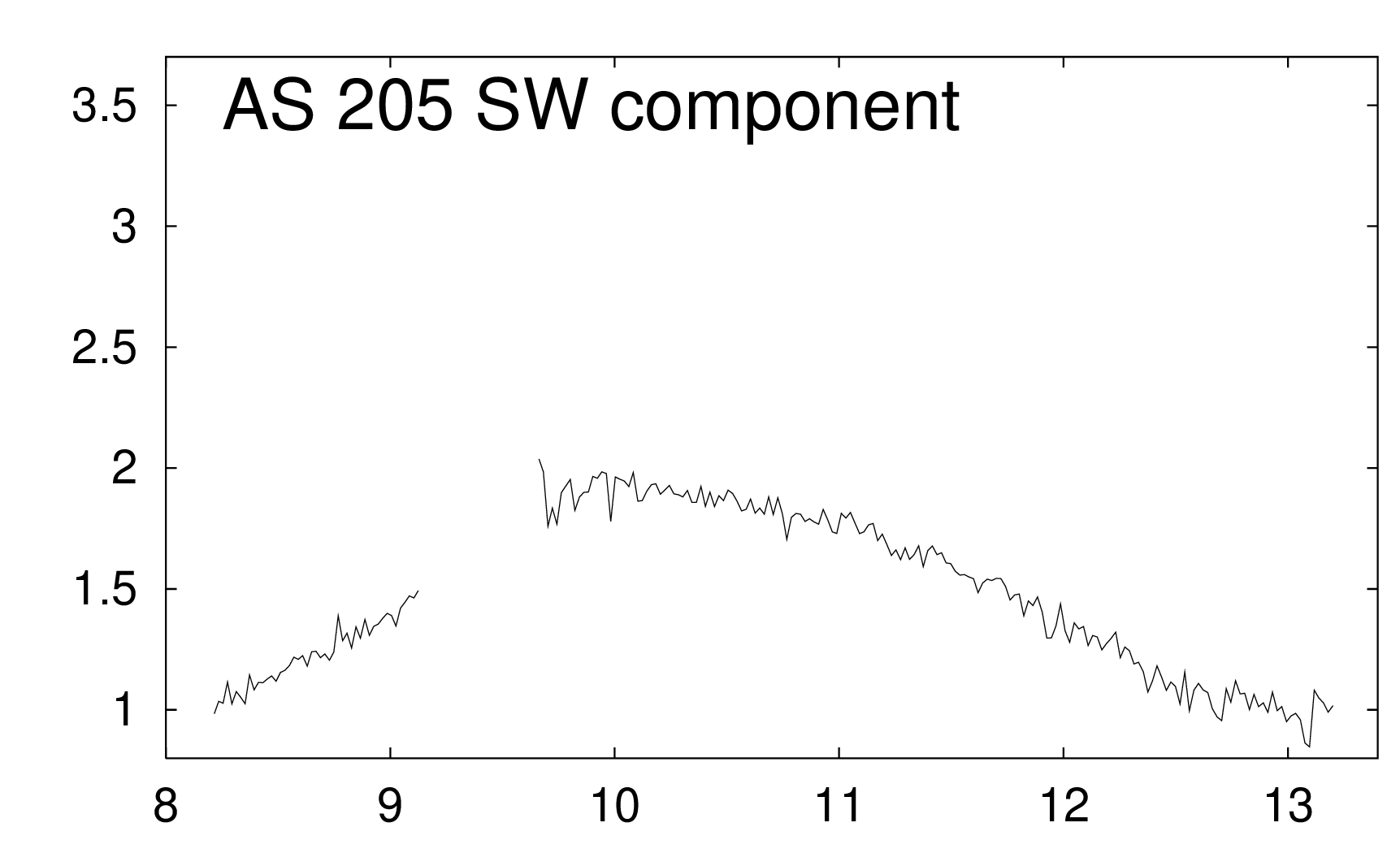,angle=270,width=4.45cm}
\epsfig{file=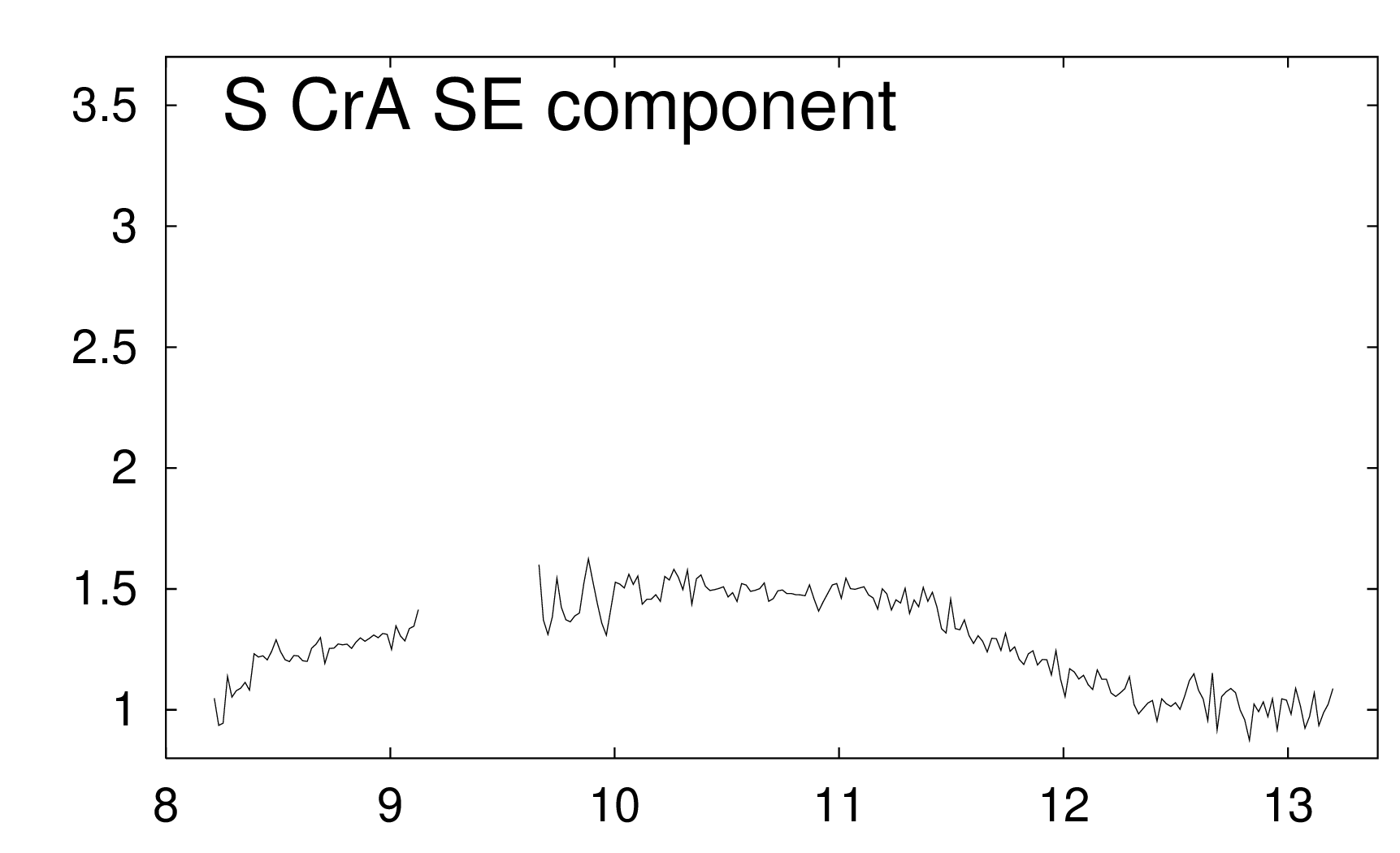,angle=270,width=4.45cm}
\epsfig{file=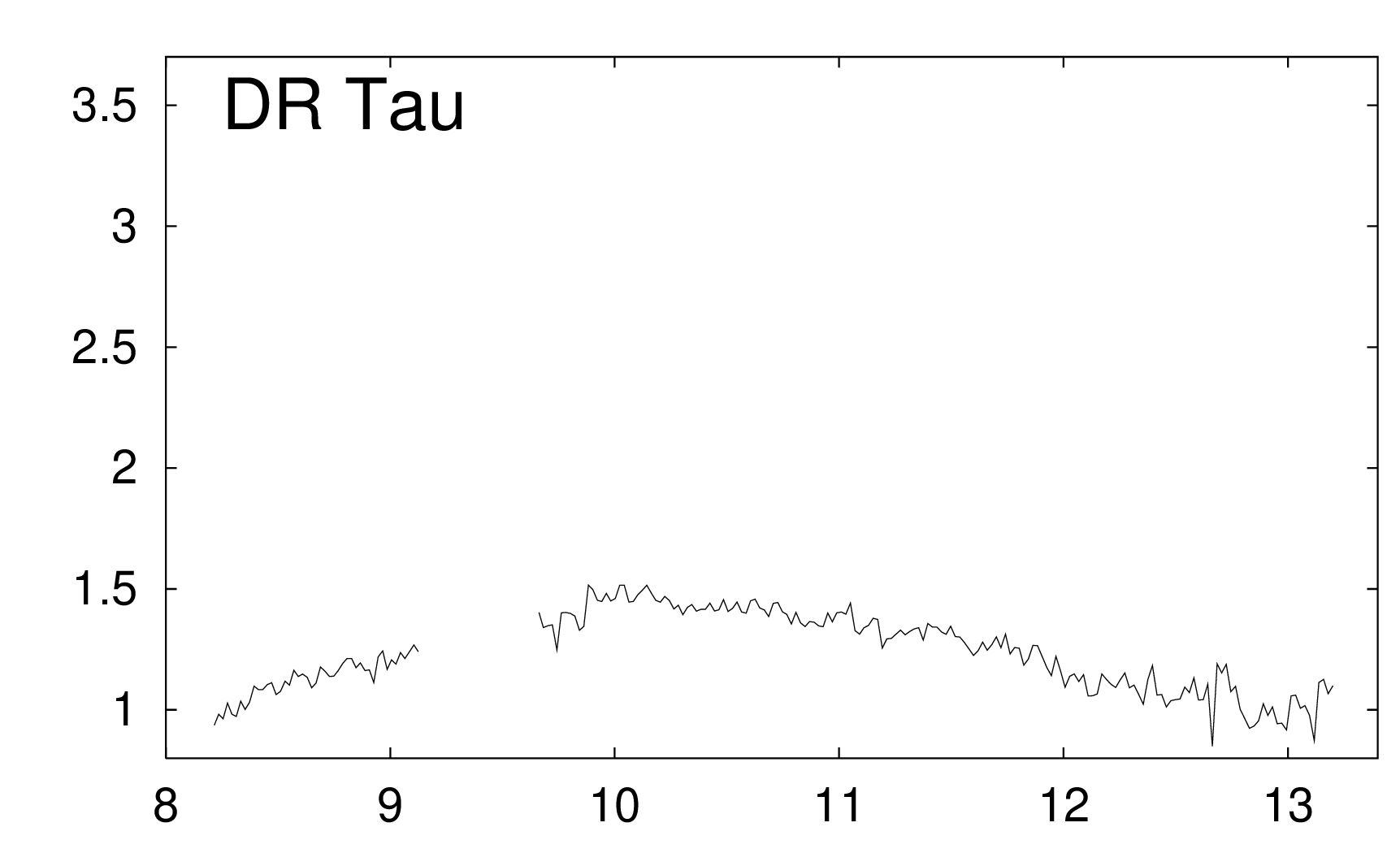,angle=270,width=4.45cm}
\epsfig{file=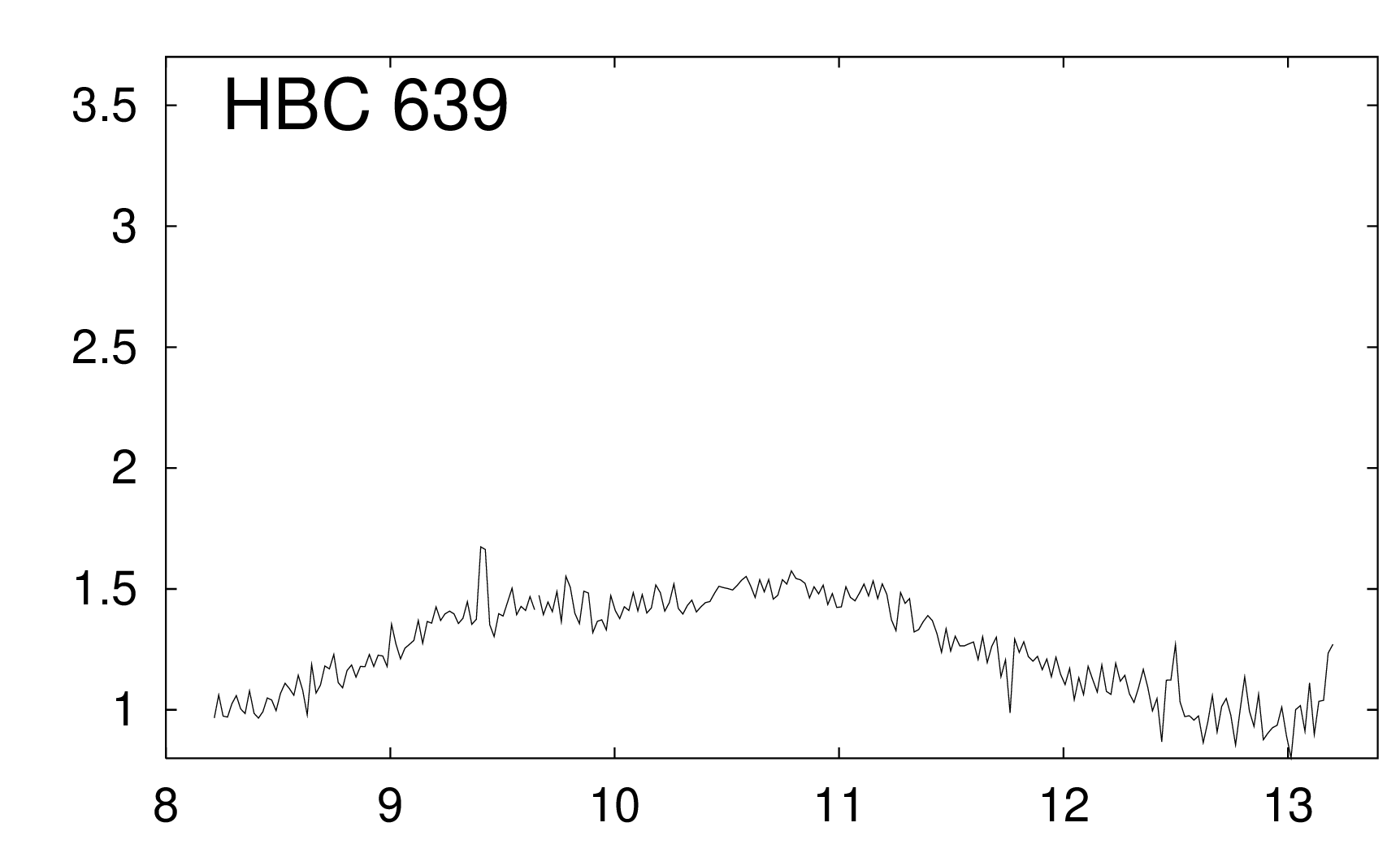,angle=270,width=4.45cm}
\epsfig{file=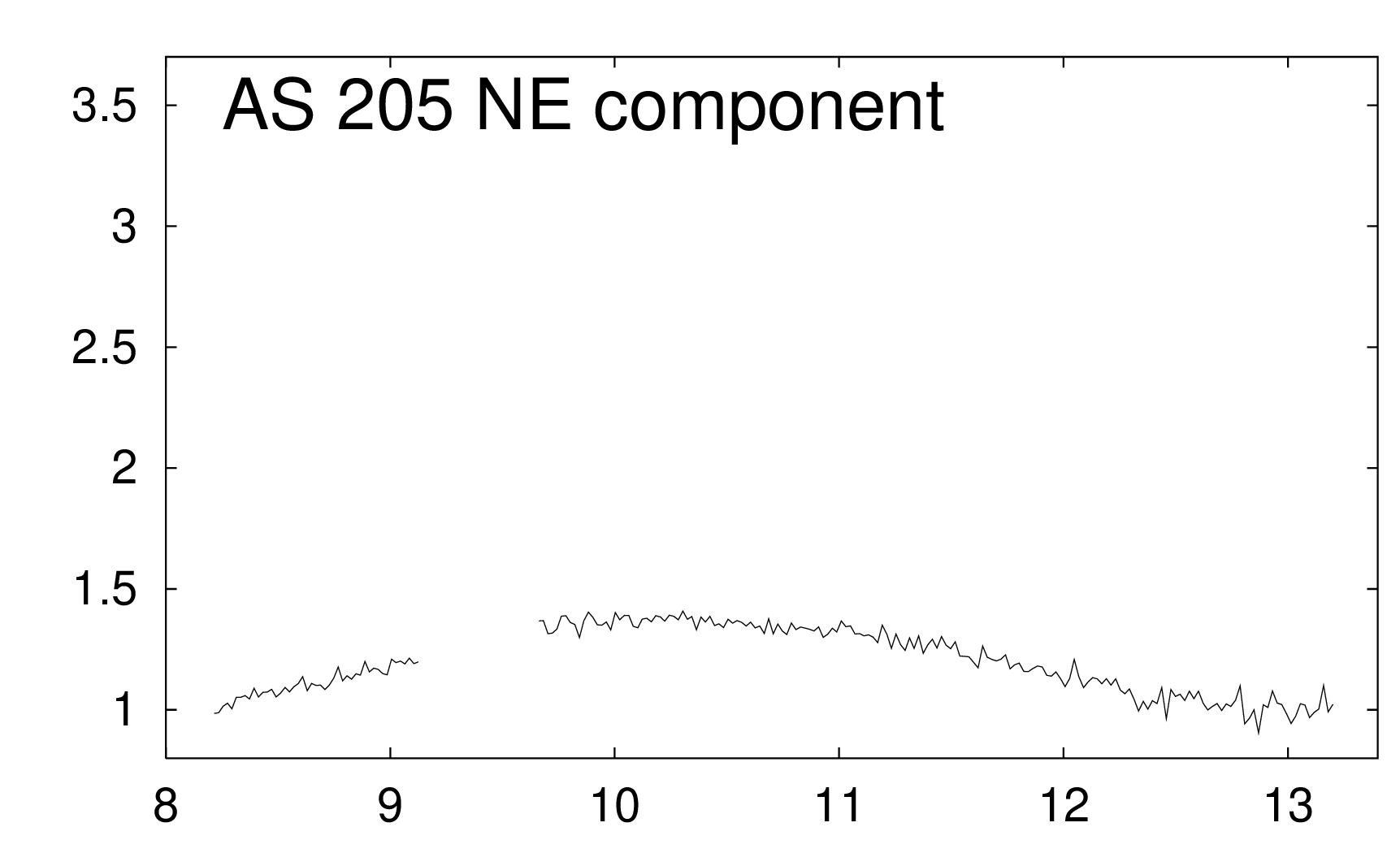,angle=270,width=4.45cm}
\epsfig{file=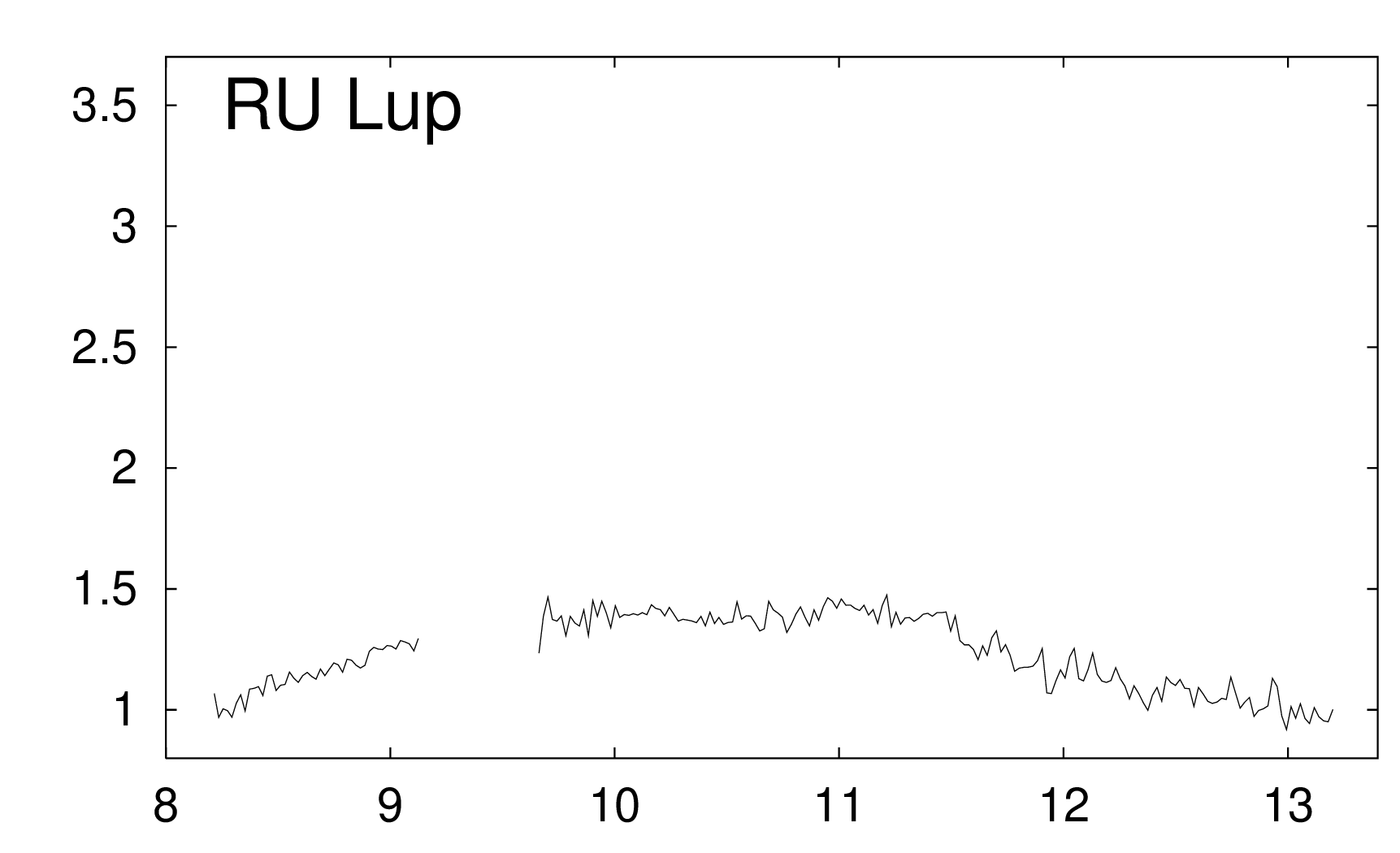,angle=270,width=4.45cm}
\epsfig{file=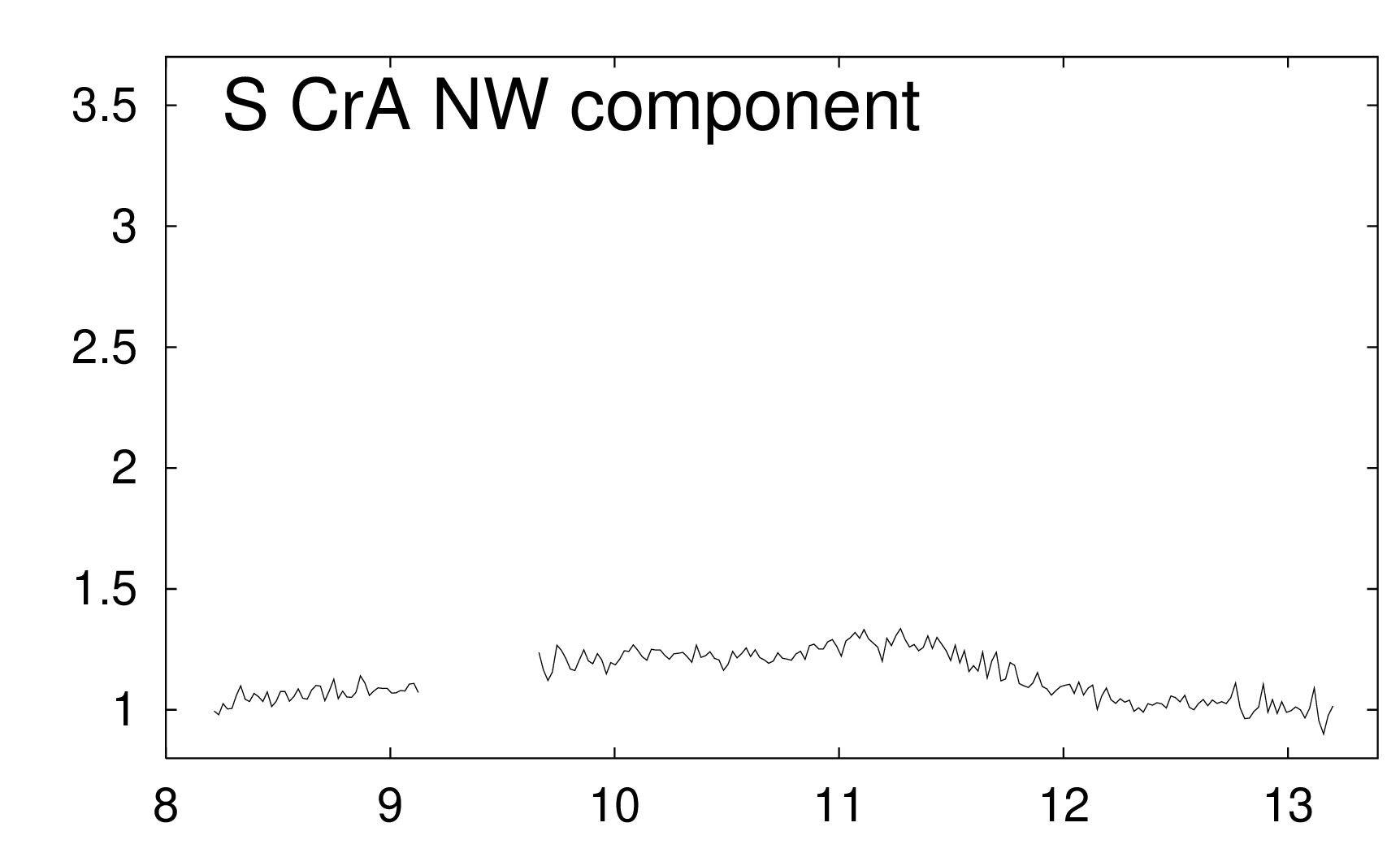,angle=270,width=4.45cm}
\vspace*{-0.3cm}
\begin{center}
\scriptsize lambda ($\mu$m)
\end{center}
\vspace*{-0.3cm}
\caption{Continuum normalized spectra of our sample ordered by the strength
  of the silicate feature. The shape of the feature is showing a correlation
  with the strength. Stronger features have a triangular shape with a 
 pronounced
  peak near 9.8\,$\mu$m while weaker features are more plateau-like. 
The gap from
  9.15 to 9.65\,$\mu$m in most of the spectra is caused by a broken channel of
  the TIMMI2 detector.}
\label{plot3}
\end{figure*}

The spectra in Fig.\,\ref{plot3} show a variety in strength and shape. In some
spectra (AK\,Sco, Haro\,1-16, CR\,Cha) a strong silicate emission with a peak
near 9.8\,$\mu$m is visible. In other cases (Ru\,Lup, DR\,Tau, HBC\,639), the
silicate emission is weaker and the profile looks more like a plateau. It is
known that the shape of the band is strongly affected by the chemical
composition and in particular by the size of the dust grains
(\cite{henning95}, \cite{bouwman01}). A correlation between strength and shape
of the silicate emission could be expected if there are systematic changes in
particle size within the group of objects.

\begin{figure}
\begin {center}
\epsfig{file=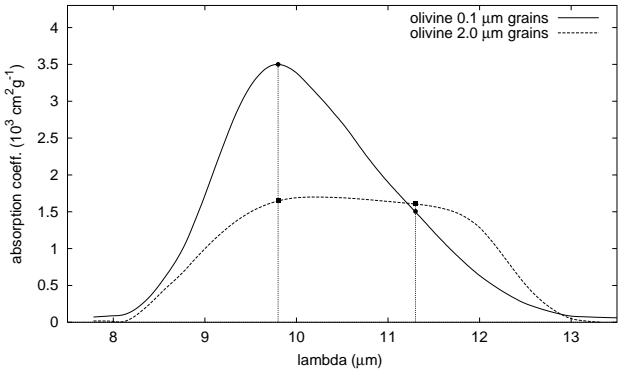,angle=270,width=8.3cm}
\end{center}
\vspace*{-0.4cm}
\caption{Theoretical spectra of olivine dust
  with grain sizes of 0.1\,$\mu$m and 2.0\,$\mu$m (from \cite{bouwman01}).
  The points at 9.8\,$\mu$m and 11.3\,$\mu$m are visualizing the dependence of
  the flux ratio at these wavelengths on the grain size.}
\vspace*{-0.3cm}
\label{plot4}
\end{figure}

To estimate the strength of the silicate emission, we use the peak of the
continuum normalised spectrum.  For the characterisation of the band shape we
measured the continuum subtracted flux at two wavelengths, 11.3 and
9.8\,$\mu$m. The flux ratio for these wavelengths can be interpreted as a
indicator for the grain size composition following Bouwman~et~al. (2001). They
point out that (see Fig.\,\ref{plot4})

\begin{itemize}
\item the mid-infrared absorption coefficient of small amorphous olivine dust
  grains (with a size of 0.1\,$\mu$m) has a triangular shape with the maximum
  at 9.8\,$\mu$m.  This spectrum is compatible with observations of the
  silicate emission of unprocessed dust of the ISM (see also \cite{bowey98}).
\item the mid-infrared absorption coefficient of larger amorphous olivine dust
  grains shows a plateau-like shape in the range from 9.5 to 12\,$\mu$m. 
\end{itemize}

Assuming a dust temperature of around 300\,K, the blackbody thermal emission
will not vary very much over our wavelength band, and the silicate emission
profiles would have a shape quite close to the shape of the absorption
coefficients.  Since the absorption cross section per unit mass of large
grains is in general smaller than that of small grains of similar shape, the
emission of large grains is in general expected to be weaker than that of
small grains.

\begin{figure}
\epsfig{file=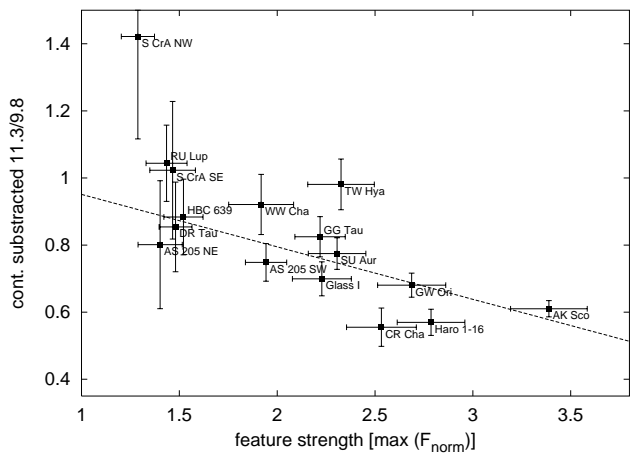,angle=270,width=8.8cm}
\caption{Correlation between the strength of the silicate feature and the
  ratio of the (continuum subtracted) fluxes at 11.3\,$\mu$m over that at 9.8
  \,$\mu$m. The strength is defined by the maximum of the silicate feature
  over the continuum.}
\label{plot5}
\end{figure}

Our observations suggest that most observed mid-infrared spectra of young low
or intermediate mass stars can be characterised simply by emission
corresponding to the two grain sizes covered in Fig.\,\ref{plot5}.  We note
that the ratio of absorption coefficients at the wavelengths of 11.3\,$\mu$m
and 9.8\,$\mu$m as function of grain size varies from 0.43 (for 0.1\,$\mu$m
grains) to about 1.0 (2.0\,$\mu$m grains). Therefore, the ratio of the
(continuum subtracted) flux at 11.3\,$\mu$m over that at 9.8\,$\mu$m should be
a measure of the grain size.  In Fig.\,\ref{plot5} we plotted this ratio
against the strength of the silicate feature. The distribution of the data
points shows a clear correlation. Strong silicate features have 11.3/9.8 flux
ratios down to 0.53 (for CR\,Cha), while weak features exhibit higher ratios
up to about 1.0 (in case of the S\,CrA NW-component even 1.42). The remaining
points are distributed over the range between these two extremes. The error
bars quoted in Fig.\,\ref{plot5} take into account the signal to noise of the
observations, and the uncertainty in the continuum estimate. A weighted linear
fit to the data points with the relation $y = a - bx$, where x is the strength
of the feature and y the flux ratio, gives the coefficients: $a=1.11\pm0.11$
and $b=-0.16\pm0.04$.  A very similar correlation between strength and shape
of the silicate emission was found by \cite{boekel03} for isolated Herbig
Ae/Be stars, with coefficients of $a=1.48\pm0.09$, $b=-0.28\pm0.04$.

Emission by Polycyclic Aromatic Hydrocarbons (PAHs) at 11.3\,$\mu$m does not
appear to be a problem for our data set.  There was no evidence for PAH
emission in our observed spectra nor in the available ISOPHOT-S spectra.  This
absence of PAH emission is not unexpected, since one can assume that the UV
radiation of T\,Tauri stars normally is too weak to cause a significant
excitation of PAH (\cite{natta95}).

\section{Discussion}

The large difference in geometric cross section per unit mass between large
and small dust grains implies that if the silicate emission band is dominated
by small grains, a substantial amount of large grains could be present without
being noticed.  On the other hand, the spectra with plateau-like emission are
in geneal weak, likely due to a much lower number of small grains present in
the upper disk regions.  In some spectra (S\,CrA both components, RU\,Lup) the
flux at 11.3\,$\mu$m is even higher than the flux at 9.8\,$\mu$m, which
results to a ratio greater than 1.  This cannot be explained by the assumption
of amorphous olivine dust with the absorption spectra shown above.  It may be
that crystalline forsterite (Mg$_2$SiO$_4$) is responsible for this effect,
since its absorption spectrum shows a peak at 11.3\,$\mu$m (see
\cite{bouwman01}, \cite{honda03}). Possible indications for crystalline
forsterite are also seen in some other objects with stronger features
(SU\,Aur, GG\,Tau).

We also note that the silicate emission emerges from warm, optically thin
material, assumed here to be located in the upper layer of a flared accretion
disk, which is irradiated by the central star.  The dust in these upper layers
of different T\,Tauri disks appears to vary continously between two extreme
types: a) small amorphous dust grains or a mixture of small and large grains,
with a spectrum very similar to the dust of the ISM and b) large amorphous
dust grains with a possible addition of crystalline silicate. If we think of
these differences in terms of evolution, then the reason for the difference
between these two dust types could be either the removal of the small grains
from the upper disk layers or the coagulation of small grains to larger ones.
A removal of small dust grains by pure settling to the mid-plane of the disk
is unlikely to explain the observed kind of differentiation since the settling
time for larger grains is shorter than for the smaller ones. A further
mechanism which could be responsible for the vanishing of small grains is by
the radiation pressure of the central star.  A star of solar luminosity would
be quite effective in removing tenth micron sized particles.  But in this
case, a permanent diffusion of particles from the mid-plane on shorter time
scales should avoid a significant decrease of the amount of small particles in
the upper layer (\cite{takeuchi03}).  The effect of coagulation is a possible
explanation, if constructive collisions to larger grains dominate over
destructive collisions which produce again small grains.  This should be the
case in T\,Tauri disks where the SED can be well fit by gas-rich disks in
hydrostatic equilibrium.

An interesting detail of our studies is the availability of separate spectra
from the components of two binary T\,Tauri systems (AS\,205 and S\,CrA). We
observed significant differences in strength and also differences in the shape
of the silicate band of the components, with the secondary having the -
relatively - stronger emission.  Since the components presumably have the same
age, the difference must have other reasons than the evolutionary state of the
stars.

\begin{acknowledgements}
This work is partly supported by the Hungarian Research Fund (OTKA
No. T037508) and by the Bolyai Fellowship (P.\`Abrah\`am).
S.Y. Melnikov thanks the German Academic Exchange Service (DAAD) for
the DAAD fellowship grant.
\end{acknowledgements}

\end{document}